\documentstyle[11pt,newpasp,twoside]{article}
\font\boldsym=cmmib10
\def    \bomega	{{\hbox{\boldsym\char'041}}}

\def\edcomment#1{\iffalse\marginpar{\raggedright\sl#1\/}\else\relax\fi}
\reversemarginpar
\begin{document}
\title{ Physics of Grain Alignment}
 \author{A. Lazarian}
\affil{Department of Astronomy, University of Wisconsin-Madison\\
%e-mail:lazarian@astro.wisc.edu
}

\begin{abstract}
Aligned grains provide one of the easiest ways to study magnetic
fields in diffuse gas and
molecular clouds. How reliable our conclusions about the inferred
magnetic field depends critically on our understanding of the physics of 
grain alignment. 
Although grain alignment is a problem of half a century standing
recent progress achieved in the field makes us believe that we are 
approaching the solution of this mystery. I review basic physical
processes involved in grain alignment
and show why mechanisms that were favored for
decades do not look so promising right now. I also discuss why the
radiative torque mechanism ignored for more than 20 years looks right now
the most powerful means of grain alignment. 

\end{abstract}

\section{Introduction}

Magnetic fields are extremely important for star formation,
galactic feedback processes etc. and polarized radiation arising from
absorption and emission by aligned grains provides
an important means for studying magnetic field topology. However,
the interpretation of polarimetry data requires  clear understanding
of processes of grain alignment and the naive rule of thumb that
dust grains are aligned everywhere and 
with longer axes perpendicular to magnetic field may be misleading
(see Goodman et al. 1995, Rao et al. 1998). 

Physics of grain alignment is deep and exciting. It is enough to say that
its study resulted in a discovery of a few new solid state effects. However,
let us start by recalling a few simple facts.
The grain alignment in interstellar medium always
 happens  in respect to magnetic field. 
It is fast Larmor
precession of grains that makes magnetic field the reference axis. Note,
that grains
 may align with their longer axes {\it perpendicular} or {\it parallel}
to magnetic field direction.
Similarly, magnetic fields may change their configuration and orientation
in space (e.g. due to Alfven waves), but if the time for such
a change is much longer than the Larmor period the alignment of
grains {\it in respect to the field lines} persists as the consequence
of preservation of the adiabatic invariant.
 
The alignment of grain axis is described by the Rayleigh reduction factor:
\begin{equation}
R\equiv \langle G(\cos^2\theta) G(\cos^2\beta)\rangle
\end{equation}
where angular brackets denote ensemble averaging, $G(x) \equiv 3/2 (x-1/3)$,
 $\theta$ is the angle between the axis of the largest moment of inertia
(henceforth the axis of maximal inertia) and the magnetic field $\bf B$, while
$\beta$ is the angle between the angular momentum $\bf J$ and $\bf B$. 
One may see (e.g. Hildebrand 1988) that $R$
is directly related to the degree of polarization. To characterize
$\bf J$ alignment in grain axes and in respect to magnetic field,
 the
measures $Q_X\equiv \langle G(\theta)\rangle$ and $Q_J\equiv \langle G(\beta)\rangle$
are used.
Unfortunately, these statistics 
are not independent and therefore $R$ is not equal to $Q_J Q_X$ (see Roberge
\& Lazarian 1999). This considerably complicates 
the treatment of grain alignment.

This review attempts to cover the recent advancements
of our understanding of grain alignment and places them in the context
of the earlier works done by giants of E. Purcell and L. Spitzer caliber.
It happened that several times the problem of grain alignment
seemed to be solved and theorists got satisfied. However, accumulation
of new observational facts and deeper insights into grain physics
caused  the changes of paradigms. Thus in what follows
we describe three periods of grain alignment theory.
A more detailed treatment of various aspects of
grain alignment the interested reader can find in earlier
reviews (e.g. Hildebrand 1988, Roberge 1996, Lazarian, Goodman \&
Myers 1997).

\section{Evolution of Grain Alignment Ideas}

{\bf Foundations.}\\
The first stage of alignment theory development
started directly after the discovery of
starlight polarization by Hiltner (1949) and Hall (1949). 
Nearly simultaneously Davis \& Greenstein (1950) 
and Gold (1951) proposed their scenarios of alignment. 

{\it Paramagnetic Alignment: Davis-Greenstein Process}\\
Davis-Greenstein
mechanism (henceforth D-G mechanism)
is based on the paramagnetic dissipation that is experienced
by a rotating grain. Paramagnetic materials contain unpaired
electrons which get oriented by the interstellar magnetic field $\bf B$. 
The orientation of spins causes
grain magnetization and the latter varies as the vector of magnetization rotates
 in grain body coordinates. This causes paramagnetic loses 
at the expense of grain rotation energy.
Note, that if the grain rotational velocity $\bomega$
is parallel to $\bf B$, the grain magnetization does not change with time
and therefore
no dissipation takes place. Thus the
paramagnetic dissipation  acts to decrease the component of $\bf \omega$
perpendicular to $\bf B$ and one may expect that eventually
grains will tend to rotate with $ \bomega \| {\bf B}$
provided that the time of relaxation $t_{D-G}$ is much shorter than  $t_{gas}$,
the
time of randomization through chaotic gaseous bombardment.
In practice, the last condition is difficult to satisfy. For $10^{-5}$ cm grains
in diffuse medium
$t_{D-G}$ is of the order of $7\times 10^{13}a_{(-5)}^2 B^{-2}_{(5)}$s , 
while  $t_{gas}$ is $3\times 10^{12}n_{(20)}T^{-1/2}_{(2)} a_{(-5)}$ s (
see table~2 in Lazarian \& Draine 1997) if
magnetic field is $5\times 10^{-6}$ G and
temperature and density of gas are $100$ K and $20$ cm$^{-3}$, respectively. 
However, in view of uncertainties in
interstellar parameters the D-G theory looked OK initially.

{\it Mechanical Alignment: Gold Process}\\
Gold mechanism is a process of mechanical alignment of grains. Consider
a needle-like grain interacting with a stream of atoms. Assuming
that collisions are inelastic, it is easy to see that every
bombarding atom deposits angular momentum $\delta {\bf J}=
m_{atom} {\bf r}\times {\bf v}_{atom}$ with the grain, 
which is directed perpendicular to both the
needle axis ${\bf r}$ and the 
 velocity of atoms ${\bf v}_{atom}$. It is obvious
that the resulting
grain angular momenta will be in the plane perpendicular to the direction of
the stream. It is also easy to see that this type of alignment will
be efficient only if the flow is supersonic\footnote{Otherwise grains
will see atoms coming not from one direction, but from a wide cone of
directions (see Lazarian 1997a) and the efficiency of 
alignment will decrease.}.
Thus the main issue with the Gold mechanism is to provide supersonic
drift of gas and grains. Gold originally proposed collisions between
clouds as the means of enabling this drift, but later papers (Davis 1955) 
showed that the process could align grains over limited patches of
interstellar space only and thus the process
cannot account for the ubiquitous grain 
alignment in diffuse medium.

{\it Quantitative Treatment and Enhanced Magnetism}\\
The first detailed analytical treatment of the problem of D-G
alignment was given by Jones \& Spitzer (1967) who described the alignment
of $\bf J$
using a Fokker-Planck equation. This 
approach allowed to account for magnetization fluctuations
within grain material and thus provided a more accurate picture of 
$\bf J$ alignment.
$Q_X$ was assumed to follow
the Maxwellian distribution, although the authors noted
that this might not be correct. 
The first numerical treatment of
D-G alignment was presented by Purcell (1969). 
By that time it became clear that the D-G
mechanism is too weak to explain the observed grain alignment. However,
Jones \& Spitzer (1969) noticed that if interstellar grains
contain superparamagnetic, ferro- or ferrimagnetic (henceforth SFM) 
inclusions, the
$t_{D-G}$ may be reduced by orders of magnitude. Since $10\%$ of
atoms in interstellar dust are iron
the formation of magnetic clusters in grains was not far fetched
(see Spitzer \& Turkey 1950, Martin 1995)
and therefore the idea was widely accepted. Indeed, with enhanced 
magnetic susceptibility the D-G mechanism was able to solve
all the contemporary problems of alignment. The conclusive
at this stage was the paper by Purcell \& Spitzer (1971) where
all various models of grain alignment, including, for
instance, the model of cosmic ray alignment by Salpeter \& Wickramasinche (1969)
and photon alignment by Harwit (1970)  
were quantitatively discussed and the D-G model with enhanced
magnetism was endorsed. It is this stage of development that is widely
reflected in many textbooks.

{\bf Facing Complexity}\\
\hspace{1in}{\it Barnett Effect and Fast Larmor Precession}\\
It was realized by Martin (1972) that rotating charged grains will develop
magnetic moment and the interaction of this moment with the interstellar
magnetic field will result in grain precession. The characteristic
time for the precession was found to be comparable with $t_{gas}$. 
However, soon  a process that
renders much larger magnetic moment was discovered (Dolginov \& Mytrophanov 
1976). This process is the 
Barnett effect, which is converse of the Einstein-Haas effect.
If in Einstein-Haas effect a paramagnetic body starts rotating
 during remagnetizations
as its flipping 
electrons transfer the angular momentum (associated with their spins)
 to the
lattice, in the Barnett effect
the rotating body shares its angular momentum with the electron
subsystem  causing magnetization. The magnetization
is directed along the grain angular velocity and the value
of the Barnett-induced magnetic moment is $\mu\approx 10^{-19}\omega_{(5)}$~erg
gauss$^{-1}$ (where $\omega_{(5)}\equiv \omega/10^5{\rm s}^{-1}$). Therefore
the Larmor precession has a period $t_{Lar}\approx 3\times 10^6 B_{(5)}^{-1}$~s and 
the magnetic field defines the axis of alignment as we explained in section~1.

{\it Suprathermal Paramagnetic Alignment: Purcell Mechanism}\\
The next step was done by Purcell(1975, 1979),
who discovered that grains can rotate much faster than were previously
thought. He noted 
that variations of photoelectric yield, the H$_2$ formation efficiency,
and  variations of accommodation coefficient over grain surface
would result in uncompensated torques acting upon a
grain. The H$_2$ formation on the grain surface clearly illustrates the
process we talk about: if H$_2$ formation takes place only over particular
catalytic sites, these sites act as miniature rocket engines
spinning up the grain. Under such uncompensated torques the grain will spin-up to
velocities much higher than Brownian and Purcell termed those
velocities ``suprathermal''. Purcell also noticed that for suprathermally
rotating grains
internal relaxation will bring $\bf J$ parallel to the axis of maximal
inertia (i.e. $Q_X=1$). Indeed, for an oblate spheroidal
grain with 
angular momentum $\bf J$  the energy can be written
\begin{equation}
E(\theta)=\frac{J^2}{I_{max}}\left(1+\sin^2\theta (h-1)\right)
\label{e}
\end{equation}
where $h=I_{max}/I_{\bot}$ is the ratio of the maximal to minimal moments
of inertia. Internal forces cannot change the angular momentum, but
it is evident from Eq.~(2) that the energy can be decreased by
aligning the axis of maximal inertia along $\bf J$, i.e. decreasing
$\theta$. Purcell (1979) discusses two possible causes of internal dissipation,
the first one related to the well known inelastic relaxation, the second is
due to the mechanism that he discovered and termed ``Barnett relaxation''.
This process may be easily understood. We know that a 
freely rotating grain preserves the direction of
$\bf J$, while angular velocity precesses about $\bf J$ and in grain body axes.
We learned earlier that the Barnett effect results in the magnetization
vector parallel to $\bomega$. As a result, the Barnett magnetization
will precess in body axes and cause paramagnetic relaxation.
The ``Barnett equivalent magnetic field'', i.e. the equivalent external
magnetic field that would cause the same magnetization of the grain  
material, is $H_{BE}=5.6 \times10^{-3} \omega_{(5)}$~G, 
which is much larger than the interstellar magnetic 
field. Therefore the Barnett relaxation happens on the scale $t_{Bar}\approx
4\times 10^7 \omega_{(5)}^{-2}$,
i.e. essentially instantly compared to $t_{gas}$ and $t_{D-G}$. 

{\it Theory of Crossovers}\\
If $Q_X=1$ and the suprathermally rotating grains are immune to randomization
by gaseous bombardment, will paramagnetic grains be perfectly aligned with
$R=1$? This question was addressed by Spitzer \& McGlynn (1979)
(henceforth
SM79) who observed
that adsorption of heavy elements on a grain  should result
in the resurfacing phenomenon that, e.g.  should remove early sites
of H$_2$ formation and create new ones. As the result,
H$_2$ torques will occasionally change their direction and spin the grain
down. SM79  showed that in the absence of
random torques the spinning down grain will
 flip over preserving the direction of its original angular momentum.
However, in the presence of random torques
 this direction will be altered with the maximal deviation inflicted
over a short period of time just before and after the flip, i.e.
during the time when the value of grain angular momentum is minimal.
The actual value of angular momentum during this critical
period depends on the ability of $\bf J$ to deviate from
the axis of maximal inertia.
SM79 observed that as the Barnett relaxation 
couples $\bf J$ with the axis of maximal inertia it 
makes randomization of grains during crossover nearly complete. With the
resurfacing time $t_{res}$ estimated by SM79 to be of the order of $t_{gas}$
the gain of the alignment efficiency was
insufficient to reconcile the theory and observations unless the grains 
had SFM inclusions.

{\it Radiative Torques}\\
If the introduction of the concept of suprathermality by Purcell changed
the way researchers thought of grain dynamics, the introduction of radiative torques
passed essentially unnoticed. Dolginov (1972) argued that quartz grains
may be spun up due to their specific rotation of polarization
while later Dolginov \& Mytrophanov (1976)
discovered that irregular grain shape may allow grains scatter left and right
hand polarized light differentially thus spinning up helical grains through
scattering of photons. They stressed that the most efficient spin-up
is expected when grains size is comparable with the wavelength and estimated
the torque efficiency for particular helical grain shapes, but failed
to provide estimates of the relative efficiency of the mechanism in the
standard interstellar conditions. In any case, this ingenious idea had not
been
appreciated for another 20 years.

{\it Observational tests: Serkowski Law}\\
All in all, by the end of seventies the the following alignment mechanisms
were known:
1. paramagnetic( a. with SMF inclusions,
   b. with suprathermal rotation),
2. mechanical,
3. radiative torques.
The third was ignored, the second was believed to be suppressed
for suprathermally rotating grains, which left
the two modifications of the paramagnetic mechanism as competing alternatives.
Mathis (1986) noticed that the interstellar polarization-wavelength dependence
known as the Serkowski law (Serkowski et al. 1975) can be explained if
grains larger that $\sim 10^{-5}$~cm are aligned, while smaller grains
are not. To account for this behavior Mathis (1986) stressed
that the SFM inclusions will have a better chance to
be in larger rather than smaller grains. The success of fitting
observational data persuaded
the researchers that the problem of grain alignment is solved at last.

{\bf New Developments}\\
Optical and near infrared observations by Goodman et al. (1992),
Goodman et al. (1995)  showed 
that polarization efficiency may
drop within dark clouds while far infrared observations by Hildebrand et al. (1984),
Hildebrand et al. (1990)
revealing aligned grains within star-forming dark clouds. This
renewed interest to grain alignment problem.

{\it New Life of Radiative Torques}\\
Probably the most dramatic change of the picture was the unexpected advent
of radiative torques. Before Bruce Draine realized that the torques
can be treated with the versatile discrete dipole approximation (DDA)
code, their role was unclear. For instance, earlier on
difficulties associated with the analytical approach to
the problem were discussed in Lazarian (1995a).
However, very soon after that Draine (1996) modified the DDA code
 to calculate the torques acting on grains of arbitrary
shape. The magnitude of torques were found to be substantial and present
for grains of various irregular shape. After that it became impossible
to ignore these torques. Being related to grain shape, rather than surface
these torques are long-lived, i.e. $t_{spin-up}\gg t_{gas}$, 
which allowed Draine \& Weingartner (1996)
to conclude that in the presence of isotropic radiation the radiative 
torques can support fast grain rotation long enough in order for
paramagnetic torques to align grains (and without any SFM
inclusions). However, the important question was what would happen
in the presence of anisotropic radiation. Indeed, in the presence
of such radiation the torques will change as the grain alignes
 and this may result in a spin-down. Moreover,
anisotropic flux of radiation will deposit angular momentum 
which is likely to overwhelm rather weak paramagnetic torques. These sort of
questions were addressed by Draine \& Weingartner (1997) and it was
found that for most of the tried grain shapes the torques tend to 
align $\bf J$ along magnetic field. The reason for that is yet unclear
and some caution is needed as the existing treatment ignores the dynamics
of crossovers which is  very important for the alignment of
suprathermally rotating grains. Nevertheless, radiative torques
are extremely appealing as their predictions are consistent
with observational data (see Lazarian, Goodman \& Myers 1995, Hildebrand et 
al. 1999). 

{\it New Elements of Crossovers}\\
Another unexpected development was a substantial change of the picture
of crossovers. As we pointed out earlier the Purcell's discovery of
fast internal dissipation resulted in a notion that $\bf J$
should always stay along the axis of maximal inertia as long
as $t_{dis}\ll t_{gas}$. Calculations in
SM79 were based on this notion.
However, this
perfect coupling
 was questioned in Lazarian (1994) (henceforth L94), where it was shown that
thermal fluctuations within grain material partially randomize
the distribution of grain axes in respect to $\bf J$. 
The process was quantified in Lazarian \& Roberge (1997)
(henceforth LR97),
where the distribution of $\theta$ for a freely
rotating grain was defined through the Boltzmann distribution
 $\exp(-E(\theta)/kT_{grain})$,
where the energy $E(\theta)$ is given by Eq.~(2). This finding
changed the understanding of crossovers a lot. First of all,
Lazarian \& Draine (1997)(henceforth LD97) observed that  thermal
fluctuations partially decouple $\bf J$ and the axis of maximal
inertia and therefore the value of angular moment at the moment
of a flip is substantially larger than SM79 assumed. Thus the
randomization during a crossover is  reduced and  LD97 obtained
a nearly
perfect alignment  for interstellar grains
rotating suprathermally, provided that
the grains were larger than a certain critical size $a_c$.  The latter 
size was found by
equating the time of the crossover and the time of the internal
dissipation $t_{dis}$. For $a<a_c$
Lazarian \& Draine (1999a) found new physical effects, which they termed
``thermal flipping'' and ``thermal trapping''. The thermal flipping
 takes place
as the time of the crossover becomes larger than $t_{dis}$.           
In this situation thermal fluctuations will enable flipovers. However,
being random, thermal fluctuations are likely to produce not a single
flipover, but multiple ones. As the grain flips back and forth the
regular (e.g. H$_2$) torques average out and the
grain can spend a lot of time rotating with thermal velocity, i.e.
being ``thermally trapped''. The paramagnetic alignment of 
grains rotating with 
thermal velocities is small (see above) 
and therefore grains with $a<a_{c}$ are
expected to be marginally aligned. The picture of preferential
alignment of large grains, as we know, corresponds to the Serkowski
law and therefore the real issue is to find the value of $a_c$.
The Barnett relaxation\footnote{A study by Lazarian \& Efroimsky (1999)
corrected the earlier estimate by Purcell (1979), but left the conclusion
about the Barnett relaxation
dominance, and therefore the value of $a_c$, intact.}
 provides a comforting value of $a_c\sim 10^{-5}$~cm. However, in a
recent paper Lazarian \& Draine (1999b) reported a new solid state effect
that they termed ``nuclear relaxation''. This is an analog of Barnett
relaxation effect that deals with nuclei. Similarly to unpaired electrons
nuclei tend to get oriented in a rotating body. However the nuclear analog
of ``Barnett equivalent'' magnetic field is much larger and Lazarian \&
Draine (1999) concluded that the nuclear relaxation can be a million times
faster than the Barnett relaxation. If this is true $a_c$ becomes of the
order $10^{-4}$~cm, which means that the majority of interstellar grains
undergo constant flipping and rotate essentially thermally in spite of
the presence of
uncompensated Purcell torques. The radiative torques
are not fixed in body coordinates and it is likely that they can provide
a means for suprathermal rotation for grains that are larger than the
wavelength of the incoming radiation. Naturally, it is of utmost importance
to incorporate the theory of crossovers into the existing codes 
and this work is under
way.

{\it New Ideas and Quantitative Theories}\\
An interest to grain alignment resulted in search of new mechanisms. For
instance, Sorrell (1995a,b) proposed a mechanism of grain spin-up due to
interaction with cosmic rays that locally heat grains and provide evaporation
of adsorbed H$_2$ molecules. However, detailed
calculations in Lazarian \& Roberge (1997b)
showed that the efficiency of the torques was overestimated; the observations
(Chrysostomou et al. 1996) did not confirm Sorrell's predictions either. 
A more promising idea that
ambipolar diffusion can align interstellar grains was
put forward in Roberge \& Hanany (1990)(calculations are done
in Roberge et al. 1995). Within this mechanism ambipolar
drift provides the supersonic velocities necessary for mechanical alignment.   
Independently L94 proposed a mechanism of mechanical grain alignment
using Alfven waves. Unlike the ambipolar diffusion, this mechanism 
operates even in ideal MHD and relies only on the difference in inertia of
atoms and grains. An additional boost to interest to mechanical
processes was gained when it was shown that suprathermally rotating
grains can be aligned mechanically (Lazarian 1995, Lazarian \& Efroimsky 1996).
As it was realized that thermally rotating grains do not $\bf J$ tightly
coupled with the axis of maximal inertia (L94) and the effect
was quantified (LR97), it got possible to formulate quantitative theories
of Gold (Lazarian 1997a) and Davis-Greenstein (Lazarian 1997b, Roberge \& 
Lazarian 1999) alignments. Together with a better understanding of
grain superparamagnetism (Draine \& Lazarian 1998) and resurfacing of
grains (Lazarian 1995c) these developments
increased the predictive power of the grain alignment theory.

{\it Alignment of PAH}\\
All the studies above dealt with classical ``large'' grains. What  about
very small (e.g. $a<10^{-7}$~cm) grains? Can they be aligned? The answer
to this question became acute after Draine \& Lazarian (1998) explained
the anomalous galactic emission in the range $10-100$~GHz as arising
from rapidly (but thermally!) 
spinning tiny grains. This rotational dipole emission will
be polarized if grains are aligned. Lazarian \& Draine (2000) (henceforth LD00)
 found
that the generally accepted picture of the D-G relaxation is incorrect
when applied to such rapidly rotating ($\omega > 10^8$~s$^{-1}$) particles. 
Indeed, the D-G mechanism assumes
that the relaxation rate is the same whether grain
rotates in stationary magnetic field or magnetic field rotates around
a stationary grain. However, as grain rotates, we know that it gets
magnetized via Barnett effect  and it is known that
the relaxation rate within a magnetized body differs
from that in unmagnetized body. A non-trivial finding in LD00
was that the Barnett magnetization provides the 
optimal conditions for the paramagnetic
relaxation which enables grain alignment at frequencies for which the D-G
process is quenched (see Draine 1996). 
LD00 termed the process ``resonance relaxation'' to
distinguish from the D-G process and calculated the expected alignment values
for grains of different sizes. Will this alignment be seen through
infrared emission of small transiently heated small grains (e.g. PAH)?
The answer is probably negative. The trouble is that internal
alignment of $\bf J$ and the axis of maximal inertia is being essentially
destroyed if a grain is heated up to high temperatures (LR97).
Therefore even if $\bf J$ vectors are well
aligned, grain axes, and therefore the direction
of polarization of emitted infrared photons, will be substantially randomized.

\section{Summary and Future work}

Let us summarize what we learned about the dynamics of grain alignment.
For a $10^{-5}$~cm grain in cold diffuse
interstellar medium the fastest motion is the grain
rotation, which happens on the time scale
less than $10^{-4}$~s. The grain tumbling and rotation of angular velocity
about $\bf J$ happens on approximately the same time scale. The alignment
of $\bf J$ with the axis of maximal inertia happens as a matter of hours
due to the very efficient nuclear relaxation. 
On the time scale of days $\bf J$ rotates about $\bf B$ due to its magnetic
moment (Dolginov \& Mytrophanov 1976), 
while gaseous damping time takes $t_{gas}\sim 10^{5}$
years. An alignment mechanism is efficient
if the alignment time is a fraction of $t_{gas}$ for
thermally rotating grains, but it may be many 
$t_{gas}$ if grains rotate suprathermally. In the latter case the dynamics
of crossovers is all-important.

At the moment radiative torques look as the most promising means of
aligning dust. Due to  thermal trapping the Purcell alignment is
suppressed. The superparamagnetic hypothesis looks OK (see Goodman \& Whittet 1996),
but the mechanism faces the problem with driving grain rotation.
The same thermal trapping makes  grain alignment
less efficient in molecular clouds where grain rotational temperature
approaches its body temperature.
It is likely that the radiative torques are still required
to drive grain rotation. 

The most challenging problem right now
is to understand the radiative torque mechanism. 
For this purpose it is necessary to describe crossovers
induced by radiative torques and include the
recently discovered flipovers into existing codes. It looks necessary to
understand why grains align (not always, but
{\it very} frequently) $\bf J$ with $\bf B$ when subjected to
anisotropic radiation. My experiments with slightly irregular
grains (using the code kindly provided to me by Bruce Draine)
 interacting with anisotropic monochromatic radiation made me believe that
it is possible to get a theoretical insight into the underlying physics.
However, whatever theory says, observational tests are necessary. Inversion of
the polarimetric data (see Kim \& Martin 1995) allows to find 
{\it for different environments} the
critical grain size starting with which grains are aligned.
Comparing this size with predictions calculated for radiative torques
should enable testing the mechanism.

Whatever the success of the radiative torques, it is necessary to proceed
with further development of alternative alignment mechanisms. Some of
them, e.g. the mechanism of mechanical alignment is suspected to cause
alignment at least in some regions (see Rao et al. 1998). Ward-Thompson et al. 
(2000) reported 850 $\mu$m polarization from 
dense pre-stellar cores, where radiative torques
should be inefficient. Could the grain larger than $a_c$ and aligned
via modified Purcell mechanism (LD97) be 
responsible? Or should we apeal to Alfven
waves or ambipolar diffusion? Further research will provide us with the answer.
In general, the variety of Astrophysical conditions
allows various mechanisms (see Lazarian, Goodman \& Myers 1997) to
have their niche. Clear understanding of grain alignment will make
polarimetry much more informative. Although so far grain alignment 
theory was applied only to interstellar environments, it is clear
that its potential is great for circumstellar and interplanetary 
studies (see Lazarian 2000).

\end{document}